\documentclass[12pt]{article}

\usepackage{epsfig}

\newcommand{\eq}{\begin{equation}}
\newcommand{\eqx}{\end{equation}}
\newcommand{\eqn}{\begin{eqnarray}}
\newcommand{\eqnx}{\end{eqnarray}}

\newcommand{\nc}{$ N_{cut} $}
\begin{document}

\title{Spectra of supersymmetric Yang-Mills quantum mechanics  }

\author{ J. Wosiek \\ \\
M.Smoluchowski Institute of Physics, Jagellonian University\\ Reymonta
4, 30-059 Krak\'{o}w, Poland}

\maketitle

\begin{abstract}
The new method of solving quantum mechanical problems is proposed. The finite, i.e. cut off, Hilbert space
is algebraically implemented in the computer code with states represented by lists of variable length.
Complete numerical solution of a given system is then automatically obtained.
The technique is applied to Wess-Zumino quantum mechanics and D=2 and D=4 supersymmetric Yang-Mills
quantum mechanics with SU(2) gauge group. Convergence with increasing cut-off was observed in many cases
well within the reach of present machines. Many old results were confirmed and some new ones, especially
for the D=4 system, are derived. Extension to D=10 is possible but computationally
 demanding for higher gauge groups.
\end{abstract}

\vspace*{1cm}
\noindent TPJU-7/02 \newline
March 2002 \newline
hep-th/0203116
\newpage
\section{Introduction}

Since the original conjecture of Banks, Fischler, Shenker and Susskind
of the equivalence between M-theory and the D=10 supersymmetric Yang-Mills quantum mechanics (SYMQM) \cite{BFSS},
a lot of effort has been put to understand, and ultimately solve, the latter \cite{CH}-\cite{GG}.

The general, not necessarily gauge, supersymmetric quantum mechanical systems have much longer history \cite{WI,CO}.
Claudson and Halpern
considered for the first time the gauge systems also well before the BFSS hypothesis \cite{CH}. In particular a complete
solution of the D=2, N=2 SYMQM was given there (see also Ref. \cite{STS}) .
Later the specific gauge models were studied in more detail and some candidates for the ground state
were constructed in the, gauge invariant, Born-Oppenheimer approximation \cite{SF,HS}.

Another important development was achieved by de Wit, L\"{u}scher
and Nicolai who have shown that the spectrum of supersymmetric
Yang-Mills quantum mechanics is continuous \cite{dWLN,NH} due to
the cancellations between fermions and bosons. This is different
from the pure Yang-Mills case where the transverse fluctuations
across the vacuum valleys do not cancel and effectively block the
valley, resulting in the discrete spectrum of the 0-volume
glueballs \cite{L}. The continuous spectrum was first regarded as
a setback, however it has turned out into a virtue with the advent
of the BFSS hypothesis and new interpretation in terms of the
scattering states. Still this new connection requires existence of
the localized state at the threshold of the spectrum - the
supergraviton \cite{POL}. This question triggered intense studies of the
Witten index of SYMQM for various D and different gauge groups
\cite{YI,ST,SM,SMK}. Powerful techniques were developed to
calculate analytically  nonabelian
integrals \cite{NSM, GG,SU} related to the Witten index. They were accompanied by
complementary and original numerical methods \cite{KS,KSII}. The
emerging picture is rather satisfactory indeed: for $D < 10$ there
is no threshold bound state while for $D=10$ there is one, exactly
as required by BFSS. This has been proven for N=2, but the
evidence is being accumulated that it holds for higher N, as well
as for other gauge groups.

The large N limit of SYMQM, which is relevant to M-theory, was studied in the framework of the mean field approximation
in Refs \cite{KAB1,KAB3}. This provided an interesting realization of the black hole thermodynamics predicted by
the M-theory \cite{POL,MAR}.

In spite of all above results, SYMQM remains unsolved for $D > 2$.
Therefore it seems natural to study this model with lattice methods which proved
so successful in treating more complex field theoretical systems.  Such a programme has been proposed in
Ref. \cite{JW}, beginning with the yet simpler (quenched, D=4, N=2) case which was later extended to
higher gauge groups $ 2<N<9 $ \cite{BW}. The asymptotic behavior in N was observed, and indications of an onset of the
interesting phase structure was found in agreement with Refs.\cite{KAB1,KAB3}.
Including dynamical fermions is possible for D=4, and may be feasible
at D=10 for the first few N. However an arbitrary N case is plagued by the sign problem caused
by the generally complex fermionic determinant.

        Another interesting approach studied by many authors,  follows the Eguchi-Kawai trick
to trade entirely the  D-dimensional configuration space into a
group space \cite{EK,TG}. Proceeding in this way one is led to
consider the fully reduced (to a single point in the Euclidean
space) model which nevertheless possesses a version of the
superymmetry \cite{IKKT}. Many analytical and numerical results
were obtained in this way (for a review see e.g. \cite{IKAMB}) .
Numerical simulations were pushed to quite high N in simplified
models \cite{AMB}. However the sign problem which is also present
there limits the Monte Carlo approach. This may be alleviated by
the new method proposed recently in Ref.\cite{AMBII}.

It is important to remember that the subject has a lot of overlap with the small volume study of gauge theories
where the valuable expertise has been accumulating for a long time
 \cite{L, LM, VABA,vBS}. Although the final goal there is to increase the space volume
and to match ultimately the standard large volume physics, the starting point is the pure Yang-Mills quantum mechanics
identical to the bosonic part of our supersymmetric systems. In fact the classical results of L\"{u}scher and M\"{u}nster
are the special case of the solution of the D=4 SYMQM in the gluino-free sector of the latter. This will be shown in Sec.5.
Recently van Baal has used the full machinery of the small volume approach to study the supersymmetric vacuum state
of the more complicated, compact version of D=4 supersymmetric YM quantum mechanics \cite{vBN}.

Summarizing, even though above quantum mechanics is much simpler that the M-theory, it remains unsolved and
thereby still poses an interesting challenge.

In this paper we propose a new approach  and present a series
 of quantitative results for simpler systems
 not always solved up to now. To this end we use the standard hamiltonian formulation of
quantum mechanics in the continuum\footnote{This work was inspired by the discussion with C. M. Bender
and the methods he developed in studying various quantum mechanical systems \cite{CMB1,CMB2}.},
construct explicitly the (finite) basis of physical states and calculate algebraically
 matrix representations of all relevant observables. This done, we proceed to calculate numerically the complete
spectrum,
the energy eigenstates, and identify (super)symmetry multiplets. This approach is entirely insensitive to
the sign problem and is equally well applicable to systems with and without fermions. Similarly to the
lattice approach, the method has an intrinsic cut-off: any quantitative results can be obtained
only within the finite dimensional subspace of the whole Hilbert space. It turns out, however, that in
all cases studied below
 many important characteristics (in particular the low energy spectrum) can be reliably obtained
before the number of basis vectors grows out of control.

In the next section we describe the method and its algebraic computer implementation. In sections 3 and 4
 the spectra of Wess-Zumino quantum mechanics and D=2 supersymmetric Yang-Mills SU(2) quantum mechanics are
derived. Section 5 contains main results of this paper. It is devoted to D=4 supersymmetric Yang-Mills
quantum mechanics with the SU(2) gauge group.
The global picture of the whole system, with the quantitative spectra in all fermionic channels and their
supersymmetric interrelations, will be presented for the first time.  Summary
 and discussion of the future applications follow in the last two Sections.

\section{Quantum mechanics in a PC}

We begin with the simple observation that the action of any quantum mechanical observable
can be efficiently implemented (e.g. in an algebraic program) if we use the discrete eigen basis
\eq
\{ |n> \}, \;\;\; |n>={1\over\sqrt{n!}}(a^\dagger)^n |0>, \label{basis}
\eqx
of the occupation number operator $a^{\dagger} a$. For example, the bosonic coordinate and momentum operators
can be written as
\eq
   x={1\over\sqrt{2}}(a+a^{\dagger}),\;\;\; p={1\over i \sqrt{2}}(a-a^{\dagger}), \label{XP}
\eqx
where all dimensionfull parameters are set to 1.
Since typical quantum observables are relatively simple functions of $x$ and $p$, they can be
represented as the multiple actions of the basic creation and annihilation
operators\footnote{The method can be also extended
to non polynomial potentials.}. Including
fermionic observables is straightforward and will be done in subsequent Sections. Generalization
to more degrees of freedom is also evident and will be carried out separately for individual systems.

The first step to quantify above considerations is to implement the Hilbert space in an algebraic program.
Any quantum state is a superposition of arbitrary number, $n_s$, of elementary states $|n>$
\eq
|st>=\Sigma_I^{n_s} a_I |n^{(I)}>, \label{stQ}
\eqx
and will be represented as a Mathematica list \footnote{Of course any other algebraic language can be used.
In more advanced
 and time consuming applications we shall be using much faster, compiler based, languages anyway.}

\eq
st=\{n_s,\{a_1,\dots,a_{n_s}\},\{n^{(1)}\},\{n^{(2)}\},\dots,\{n^{(n_s)}\}\}, \label{stM}
\eqx
with $n_s+2$ elements. The first element specifies the number of elementary states entering the linear combination,
Eq.(\ref{stQ}), the second is the sublist supplying all (in general complex) coefficients $a_I, I=1,\dots,n_s $,
and the remaining $n_s$ sublists specify the occupation numbers of elementary, basis states.
In particular, convention (\ref{stM}) implies for an elementary state
\eq
|n> \leftrightarrow  \{1,\{1\},\{n\} \}.
\eqx
In the case of more degrees
of freedom the single occupation numbers $n^{(I)}$ in Eq.(\ref{stM}) will become lists themselves containing
all occupation numbers corresponding to the individual degrees of freedom.

In the next step we implement basic operations defined in the Hilbert space: addition of two states,
multiplication by a number and the scalar product. All these can be simply programmed as definite
operations on Mathematica lists. Table 1 displays corresponding routines which transform lists in accord
with the principles of quantum mechanics. This done, it is a simple matter to define the creation and annihilation
operators which act as a list-valued functions on above lists. This also defines the action of the position
and momentum operators according to Eq.(\ref{XP}). Then we proceed to define any quantum observables of interest:
hamiltonian, angular momentum, generators of gauge transformations and supersymmetry generators, to name only the
most important examples.

 \begin{table}
  \begin{center}
   \begin{tabular}{cccc} \hline\hline
   operation  &  quantum mechanics        &   PC  &  mapping \\
\hline\hline

  any state   & $ |st> $            &      $list$               &                                         \\
  sum         & $ |st_1>+|st_2> $ &  $ add[list_1,list_2]$  &  $ (list_1,list_2)\rightarrow list_3  $ \\
  number multiply  & $ \alpha |st_1> $ &  $ mult[\alpha,list_1]$ &  $ list_1 \rightarrow list_2    $       \\
  scalar product & $ <st_1|st_2> $ &  $ sc[list_1,list_2]$ &  $ (list_1,list_2)\rightarrow number  $ \\
  empty state   & $ |0> $            &      \{1,\{1\},\{0\}\}               &                                         \\
    null state   & $ 0 $            &      \{0,\{\}\}               &                                         \\

   \hline\hline
   \end{tabular}
  \end{center}
\caption{ Quantum states in the Hilbert space and their computer implementation.}
\end{table}

Now our strategy should be obvious: given a particular system, we define the list corresponding to the empty state,
then we generate a finite basis of $N_{cut}$ vectors and calculate matrix representations of the hamiltonian
and other quantum operators using above rules. Thereby the problem is reduced to a simple question in linear algebra,
namely the spectrum of the system is given by the eigenvalues and eigenvectors of the hamiltonian matrix,
their behaviour under rotations by the angular momentum matrix, etc.

Of course the most important question is how much the ultimate
physics at $N_{cut}=\infty$ is distorted by the finite cut-off. 
This can be answered quantitatively by inspecting the dependence
of our results on $N_{cut}$ for practically available sizes of the bases. In all systems studied so far
(and discussed below) the answer is positive: one can extract the meaningful (i.e. $N_{cut}=\infty$) results
before the size of the basis becomes unmanageable. However this can be answered only {\em a posteriori}
and individually for every system.

The method outlined above turns out to be a rather powerful tool capable to solve quantitatively various
quantum mechanical problems with finite but large  number of degrees of freedom.

\section{Wess-Zumino quantum mechanics}

   This system is obtained by dimensional reduction of the four-dimensional Wess-Zumino model leaving only
the time dependence of the dynamical degrees of freedom. Resulting quantum mechanics (WZQM) has one complex
bosonic variable
$\phi(t)=x(t)+i y(t)\equiv x_1(t)+i x_2(t)$ and two complex Grassmann-valued fermions
$\psi_{\alpha}(t),\;\;\alpha=1,2$ \cite{SHIF}.
The hamiltonian reads
\eq
H={1\over 2}(p_x^2+p_y^2)+|F|^2+\left( (m+2g\phi)\psi_1\psi_2+h.c.\right),  \label{WZham}
\eqx
where the auxiliary field $F$ is determined in terms of $ \phi $ by the classical equation of motion,
\eq
F=-(m\phi + g\phi^2),
\eqx
since it enters only quadratically into the action.
Independent variables satisfy the standard commutation relations
\eq
[x_i,p_k]=i\delta_{ik} ,\;\; \{\psi_{\alpha},\psi_{\beta}^{\dagger}\}=\delta_{\alpha\beta},
\eqx
which allow us to parameterize these coordinates in terms of the creation and annihilation operators
\eqn
   x={1\over\sqrt{2}}(a_x+a_x^{\dagger}),\;\;\; p_x={1\over i \sqrt{2}}(a_x-a_x^{\dagger}),  \\
   y={1\over\sqrt{2}}(a_y+a_y^{\dagger}),\;\;\; p_y={1\over i \sqrt{2}}(a_y-a_y^{\dagger}),  \\ \label{XFP}
   \psi_{\alpha}=f_{\alpha},\;\;\psi_{\alpha}^{\dagger}=f_{\alpha}^{\dagger},
\eqnx
which in turn satisfy
\eq
[a_i,a_k^{\dagger}]=\delta_{ik},\;\;\;\{f_{\alpha},f_{\beta}^{\dagger}\}=\delta_{\alpha\beta}.  \label{crWZ}
\eqx
Now we implement this quantum mechanical system in the computer as explained in the previous Section.
The empty state will be represented by a list
\eq
|(0,0),(0,0)> \leftrightarrow \{1,\{1\},\{\{0,0\},\{0,0\}\}, \label{wzvac}
\eqx
where the first brackets specify  occupation numbers of the bosonic, and the second fermionic, states.
To define properly the fermionic creation/annihilation operators we follow the original construction
of Jordan and Wigner \cite{JOWI, BD}
\eq
f_{\alpha}=\Pi_{\beta<\alpha}(-1)^{F_{\beta}}\sigma_{\alpha}^{-}\;\;\;
f_{\alpha}^{\dagger}=\Pi_{\beta<\alpha}(-1)^{F_{\beta}}\sigma_{\alpha}^{+},
\eqx
where the fermionic number operator $F_{\alpha}= f_{\alpha}^{\dagger} f_{\alpha}$ (no sum) and $\sigma^{\pm}_{\alpha}$ are rising and lowering operators,
commuting for different $\alpha$, with $(\sigma_{\alpha}^{\pm})^2=0$ . The Jordan-Wigner phases ensure uniform anticommutation
rules for fermionic operators.

We proceed to construct the finite eigen basis of the fermionic, $F_{\alpha}$, and bosonic, $B_i$, number operators.
For more than one degree of freedom the organization of a basis and definition of the cut-off, $N_{cut}$, is not unique.
In this case we have chosen the cut-off to be the maximal number of bosonic quanta. That is, the basis subject to the
cut-off \nc\ consists of all orthonormal elementary states with $B=B_1+B_2 \le N_{cut}$ and all allowed fermionic quanta,
i.e. $F=F_1+F_2\le 2 $. In practice the basis is created by action on the empty state with all elementary independent monomials
of bosonic creation operators up to $N_{cut}$'th order, followed by the three independent monomials of fermionic creation
 operators. Hence the Hilbert space, subject to the cut-off $N_{cut}$, has $2(N_{cut}+1)(N_{cut}+2)$ dimensions.

Given the basis, the matrix representation of the hamiltonian (and any other observable of interest) can be readily
calculated with our computer based "quantum algebra". The spectrum and energy eigenstates are then obtained
by numerical diagonalization of the hamiltonian matrix.

\begin{figure}[htb]
\epsfig{width=12cm,file=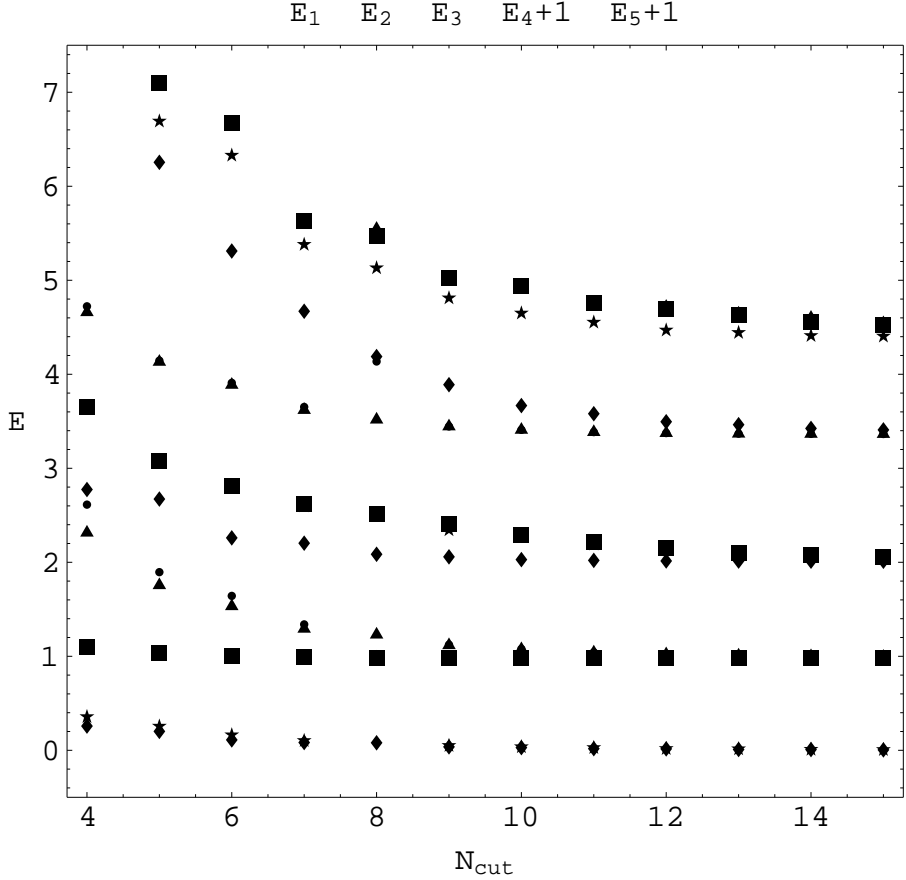}
\caption{ Spectrum of WZ quantum mechanics as a function of cut-off, \nc . }
\label{f1}
\end{figure}

Figure \ref{f1} shows the spectrum of low energy states as a function of the cut-off up to \nc=15 for $m=g=1$.
Energies of the fourth and fifth level are shifted by 1 to avoid confusion of levels for low \nc.
Clear convergence with \nc\ is seen, well within a capacity of a medium class PC.  The convergence
is faster for lower states. This general feature of our method can be easily understood. The basis generated
by creation operators, Eq.(\ref{crWZ}), is nothing but the eigen basis of some normalized harmonic oscillator.
It is obvious that the ground state can be easier approximated by a series of harmonic oscillator wave functions
that the higher states with all their detailed structures, zeroes, etc. Quantitatively, at \nc=10 the exact supersymmetric
ground state energy is reproduced with the 3\% accuracy which further improves to below 1\% at \nc=15
\footnote{Since the exact value is zero,  we take the first excited energy as a reference scale.}.
 Supersymmetric pattern of the spectrum is also evident. First, the lowest bosonic state has no
fermionic counterpart, its energy tends to zero, hence we identify it with the supersymmetric vacuum of the model
\footnote{The model exhibits additional two-fold degeneracy in bosonic and fermionic sectors which is not related to
supersymmetry \cite{SHIF}}. Second, all higher states group into bosonic-fermionic multiplets with the same energy.
Splittings inside the multiplets decrease with \nc\ as shown in Table 2.
We conclude that supersymmetry at low energies is restored at the level of few percent in the cut Hilbert space
 with states containing up to 15 bosonic quanta.

\begin{table}
  \begin{center}
   \begin{tabular}{cccc} \hline\hline
 $N_{cut}$ &  $ M_1 $   & $ M_2 $   & $ M_3 $     \\  \hline
  9        &  0.1321  & 0.1574  & 0.1835    \\
  10       &  0.0924  & 0.1213  & 0.1245   \\
  11       &  0.0595  & 0.0941  & 0.0886  \\
  12       &  0.0420  & 0.0657  & 0.0609   \\
  13       &  0.0282  & 0.0440  & 0.0407  \\
  14       &  0.0212  & 0.0332  & 0.0287  \\
  15       &  0.0149  & 0.0207  & 0.0192  \\

   \hline\hline
   \end{tabular}
  \end{center}
\caption{Splittings within the first three supersymmetric multiplets for different cut-offs.}
\end{table}

Let us see how the above convergence to the supersymmetric spectrum shows up in the Witten index,
which in this approach can be calculated directly from the definition
\eq
I_W(T)=\Sigma_{i} (-1)^{F_i}\exp{(-T E_i)}.
\eqx

\begin{figure}[htb]
\epsfig{width=12cm,file=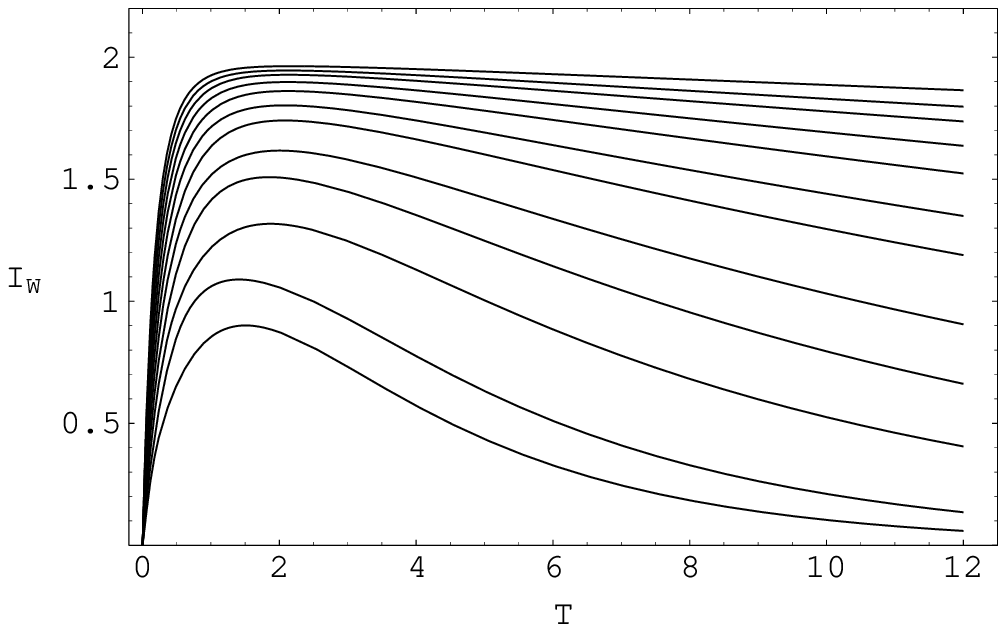}
\caption{ Witten index for Wess-Zumino quantum mechanics. }
\label{f2}
\end{figure}

Fig.\ref{f2} shows Witten index for $ 4 < N_{cut} < 15 $. Nice convergence to the exact, time independent,
result $I_W(T)=2$ is observed. Two things are happening which ensure this behaviour. First, there
are two (due to the above degeneracy within bosonic and fermionic sectors) supersymmetric
ground states. At the largest cut-off, \nc=15, their energies are not exactly zero but tiny
$E_{0,0}=0.0046,\;E_{0,1}=0.0071$, hence giving rise to the exponential fall-off at large $T$ with
a very small slope. Second, cancellations between non zero energy states within supersymmetric
multiplets are not exact and still leave exponential terms, albeit with smaller and smaller
(with increasing \nc ) coefficients. This gives the residual time dependence which is however
much weaker than the exponential one.

Another, rather general feature of $I_W(T)$, can be also observed here. Exactly at $T=0$
Witten index vanishes for every \nc\ just because the bases generated with our procedure
 automatically satisfy a "global" supersymmetry requirement. Namely the number of bosonic
 and fermionic states
in a basis is the same for every \nc . Therefore we conclude that
\eq
\lim_{N_{cut}\rightarrow\infty} I_W(0,N_{cut})=0.
\eqx
On the other hand exact supersymmetry requires that this complete correspondence between
fermionic and bosonic states is violated since the ground state at $E=0$ does not have its fermionic
counterpart while each non zero energy state has. This apparent paradox is reconciled by noting that
exact supersymmetry emerges only in the limit of the infinite \nc . In another words the unbalanced
fermionic state is pushed (with increasing \nc) to higher and higher energies and eventually
 "vanishes from the spectrum".
However it leaves a visible effect - Witten index has a discontinuity at $T=0$
which can be expressed as the  noncommutativity of the
$T\rightarrow 0$ and $N_{cut}\rightarrow\infty$ limits
\eq
\lim_{N_{cut}\rightarrow\infty} \lim_{T\rightarrow 0}I_W(T,N_{cut})\ne
\lim_{T\rightarrow 0}\lim_{N_{cut}\rightarrow\infty} I_W(T,N_{cut}).
\eqx
The right hand side of the above equation, known as the bulk contribution,
may not be integer for the continuous spectrum.

Above results were obtained for $m=g=1$, but the method works equally well for any other choices of parameters.
The case $m=0$ has a rotational symmetry and can be solved by other methods \cite{FromR}.
To our knowledge no quantitative calculation of the spectra and wave functions for $m\ne 0$ exists.
As mentioned earlier this approach provides a complete quantum mechanics of a system. In particular,
wave functions in the coordinate representation are simply given by the well defined linear combinations of the
harmonic oscillator wave functions.
More detailed discussion of the model, and comparison with other authors will be presented elsewhere.

\section{D=2, supersymmetric, SU(2) Yang-Mills quantum mechanics}

   This system, although solved analytically\cite{CH}, has two new features: gauge invariance and continuous
spectrum.
Therefore we have chosen it as a next exercise. Reducing from $D=2$ to one time dimension one is left
with the three real, colored, bosonic variables $x_a(t)$ and three complex, fermionic degrees
of freedom $\psi_a(t)$ also in the adjoint representation of SU(2) , $a=1,2,3$.

The hamiltonian reads \cite{CH}
\eq
H={1\over 2} p_a p_a +i g \epsilon_{abc} \psi_a^{\dagger} x_b \psi_c, \label{HD2}
\eqx
where the quantum operators $x,p,\psi,\psi^{\dagger}$ satisfy
\eq
[x_a,p_b]=i\delta_{ab} ,\;\; \{\psi_a,\psi_b^{\dagger}\}=\delta_{ab}.
\eqx
Hence they can be written in terms of the creation and annihilation operators as before
\eqn
   x_a={1\over\sqrt{2}}(a_a+a_a^{\dagger}), & p_a={1\over i \sqrt{2}}(a_a-a_a^{\dagger}),  \\ \label{XPD2}
   \psi_a=f_a, & \psi_a^{\dagger}=f_a^{\dagger}.
\eqnx
To implement fermionic creation and annihilation operators we again used the Jordan-Wigner construction.

The system has a local gauge invariance with the generators
\eq
G_a=\epsilon_{abc}(x_b p_c - i \psi_b^{\dagger} \psi_c). \label{GD2}
\eqx
Therefore the physical Hilbert space consists only of the gauge invariant states.
This can be easily accommodated in our scheme noting that the gauge generators of the SU(2)
are just the angular momentum operators acting in color space. In fact the fermionic part of
Eq.(\ref{GD2}) can be also interpreted in this way since the momentum canonically conjugate to
$\psi$: $\pi_{\psi}=i\psi^{\dagger}$. Therefore we construct all possible invariant under SU(2) combinations
of the creation operators (referred for short as creators) and use them to generate a complete
gauge invariant basis of states. There are four lower order creators:
\eqn
a_a^{\dagger} a_a^{\dagger}\equiv (aa),\\ \nonumber
a_a^{\dagger} f_a^{\dagger}\equiv (af),\\ \nonumber
\epsilon_{abc}a_a^{\dagger} f_b^{\dagger}f_c^{\dagger} \equiv (aff),\\ \label{creators}
\epsilon_{abc}f_a^{\dagger} f_b^{\dagger}f_c^{\dagger} \equiv (fff).\\  \nonumber
\eqnx
Pauli principle implies that $ (ff)=(af)^2=(aff)^2=(fff)^2=0 $, therefore the whole basis can be conveniently
organized into the four towers of states, each tower beginning with one of the following states
\eq
|0_F>=|0>;\;\; |1_F>=(af)|0>,\;\; |2_F>=(aff)|0>,\;\; |3_F>=(fff)|0>, \label{ground}
\eqx
where we have labeled the states by the gauge invariant fermionic number $F=f_a^{\dagger}f_a$.
The empty state (and its Mathematica representation) reads
\eq
|(0,0,0),(0,0,0)> \leftrightarrow \{1,\{1\},\{\{0,0,0\},\{0,0,0\}\}\}, \label{D2vac}
\eqx
with the obvious assignment of bosonic and fermionic occupation numbers. To obtain the whole basis
it is now sufficient to repeatedly act on the four vectors, Eq.(\ref{ground}), with the bosonic creator (aa), since
action with other creators either gives zero, due to the Pauli blocking, or produces linearly dependent state
from another tower. Basis with the cut-off \nc\ is then obtained by applying $(aa)$ up to \nc\ times to each of the four
"ground" states. Of course our cut-off is gauge invariant, since it is defined in terms of the gauge invariant creators.
Moreover, since the gauge generators can be implemented into the algebraic program as any other observables,
we can directly verify gauge invariance of the basis and any other state in question.

    Once the above basis is constructed, we proceed to calculate the matrix
representation of the hamiltonian, the spectrum and the eigenstates according to Sec.2
\footnote{All numerical results are for $g=1$.}. Again all computation can be done on
a small PC and the longest run, corresponding to 20 bosonic quanta, takes approximately 500 sec. It is worth
to mention that the most time consuming are the algebraic operations on the states (c.f. Table 1) in the
PC-based abstract Hilbert space. Consequently the program spends most of the time calculating the matrix
representations of various operators \footnote{In fact this part can be (and often is)
done analytically in the integer arithmetic.} while the numerical diagonalization is rather fast.

    The hamiltonian, Eq.(\ref{HD2}) can be rewritten
\eq
H={1\over 2} p_a p_a + g x_a G_a, \label{HD2free}
\eqx
hence it reduces to that of a free particle in the physical, gauge invariant basis. It follows that it preserves
the gauge invariant fermionic number and, consequently, can be diagonalized independently
in each sector spanned by the four towers in Eq.(\ref{ground}).

\begin{figure}[htb]
\epsfig{width=12cm,file=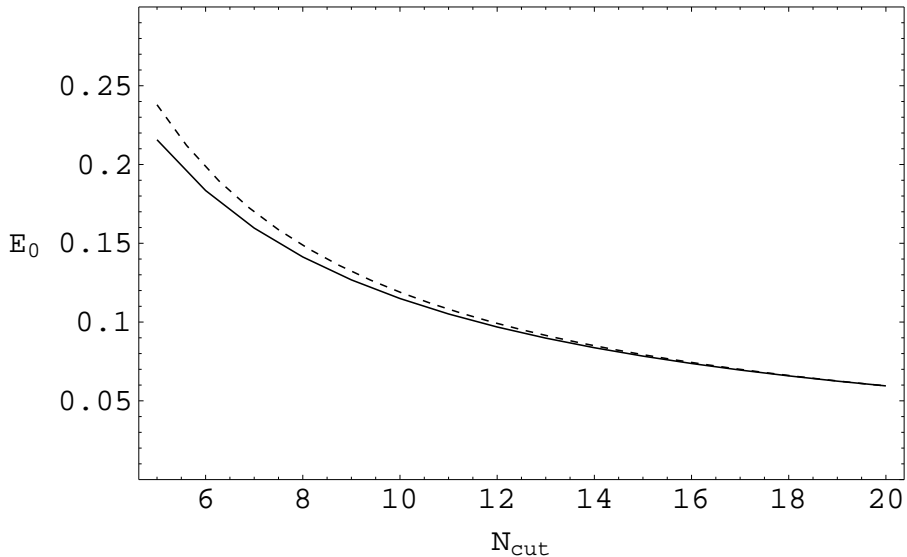}
\caption{ D=2 supersymmetric Yang-Mills quantum mechanics.
The energy of the lowest state as a function of he cut-off and the 1/\nc\ fit (dotted line).}
\label{f3}
\end{figure}

The spectrum is doubly degenerate because of the particle-hole symmetry which relates empty and filled
fermionic states ($ |0_F> \leftrightarrow |3_F> $) and their 1-particle 1-hole counterparts
($ |1_F> \leftrightarrow |2_F> $). This symmetry is not violated by the cut-off and indeed we see it exactly
in the spectrum. On the other hand, supersymmetry
connects sectors which differ by 1 in the fermionic number e.g. ( $|0_F> \leftrightarrow  |1_F>$ ) and in general
is restored
only at infinite \nc\ . A good measure of the SUSY violation is provided by the energy of the
ground state which should be 0.  We see that the energy of the lowest (doubly degenerate)
state converges to 0 but rather slowly, c.f. Fig.\ref{f3}. This is interpreted as the indication that the
spectrum is continuous
at infinite cut-off. Indeed it is hard to approximate the non localized state by the harmonic oscillator
states (this in fact is our basis) and consequently the convergence of such an approximation must be slow
\footnote{This argument can be turned into a proof that the free spectrum converges like 1/\nc
\cite{TW} which is also confirmed here (c.f. Fig.\ref{f3}).}. This deficiency can be sometimes turned into
 an advantage as will be seen in the D=4 case.

     Surprisingly however, there exists a particular scheme of increasing the basis which renders {\em exact}
supersymmetry at every finite cut-off \nc\ in this model. To see that, let us first discuss tests of the SUSY on the
operator level. SUSY generators read
\eq
Q=\psi_a p_a,\;\;\;\bar{Q}_a=\bar{\psi}_a p_a,
\eqx
with $\bar{\psi}=\psi^{\dagger}$, and satisfy the algebra
\eq
\{Q,\bar{Q}\}=2 H-g x_a G_a,\;\;Q^2=\bar{Q}^2=0.   \label{SUSYALGD2}
\eqx
Since the matrix elements of the SUSY generators can be easily calculated in our PC-based Hilbert space approach,
one can readily check the matrix element version of Eqs.(\ref{SUSYALGD2}). Indeed we find that these relations
are satisfied to better and better precision with increasing \nc . If so, it is natural to ask what is the spectrum
of the hamiltonian matrix $H^{Q\bar{Q}}$  {\em defined} by Eq.(\ref{SUSYALGD2})  at finite cut-off.
To this end we diagonalize the matrix
\eq
H^{Q\bar{Q}}_{N,N'}={1\over 2}(<N|Q|M><M|\bar{Q}|N'>+<N|\bar{Q}|M><M|Q|N'>),
\eqx
where $N,N'$ and $M$ label vectors of our finite basis\footnote{The second term in
Eq.(\ref{SUSYALGD2}) does not contribute in the physical basis.}. Of course SUSY generators mix the fermionic number, therefore
one should combine all four towers of Eq.(\ref{ground}) into one big basis. One more trick is required to achieve exact SUSY
at finite cut-off: we have to increase the basis allowing from the beginning for the disparity between the fermionic
and bosonic sectors. That is, the size of the basis grows as: $2+4+4+\dots$. Hence dimension of the
cut Hilbert space is $2+4 N_{cut}$ in this scheme. With this choice the spectrum of the $ H^{Q\bar{Q}} $
has exact superymmetry as shown in Table 3.

 \begin{table}
  \begin{center}
   \begin{tabular}{ccccccc} \hline\hline
   $2+4N_{cut}$   & $ 10 $ & $ 14 $ & $ 18 $ &  22  &  $26$  & g \\
\hline\hline
 $ E_0 $      &  0     &  0     & 0      &   0   &  0     & 2 \\
 $ E_1 $      &  0.815 &  0.610 & 0.489  & 0.409 & 0.351  & 4 \\
 $ E_2 $      &  2.685 &  1.904 & 1.495  & 1.236 & 1.056  & 4 \\
 $ E_3 $      &        &  4.235 & 3.160  & 2.558 & 2.160  & 4 \\
 $ E_4 $      &        &  0     & 5.856  & 4.522 & 3.743  & 4 \\
 $ E_5 $      &        &  0     &        & 7.525 & 5.960  & 4 \\
 $ E_6 $      &        &  0     &        &   0   & 9.230  & 4 \\
   \hline\hline
   \end{tabular}
  \end{center}
\caption{The spectrum and the degeneracy, $g$, of the hamiltonian
$H^{Q\bar{Q}} $ defined at finite \nc\  as $\{Q,\bar{Q}\}/2$. }
\end{table}
As required, the ground state has zero energy and does not have the supersymmetric image while higher states
form supersymmetric doublets with the same energy. Additional degeneracy is caused by the particle-hole
symmetry as explained earlier. All non zero eigenvalues (with finite "principal" quantum number) tend to zero
with increasing \nc\ and form, at $N_{cut}=\infty$, the continuum spectrum of a free hamiltonian, Eq.(\ref{HD2free}).
To reach the solutions with non zero energy, one should appropriately scale the index of an eigenvalue with \nc\ .
The four towers of eigenstates are in the direct correspondence with the four gauge invariant plane waves
constructed in Ref.\cite{CH}.

To our knowledge this is the first case where the cut-off preserves exactly the supersymmetry.
On the other hand this success of the particular cut-off scheme may be related to the simple structure
of the D=2 SYM quantum mechanics and its complete solubility. Nevertheless, it is important to keep in mind
that at finite \nc\ one is free to define the hamiltonian within the "O(1/\nc)" tolerance, i.e. as long as
various definitions converge to the same limit at infinite \nc . We decided to use $H^{Q\bar{Q}}$
because it is the anticommutator of charges which is used to derive basic properties of the SUSY spectrum.
Further study of this exact realization of supersymmetry  will be continued elsewhere.

\begin{figure}[htb]
\epsfig{width=12cm,file=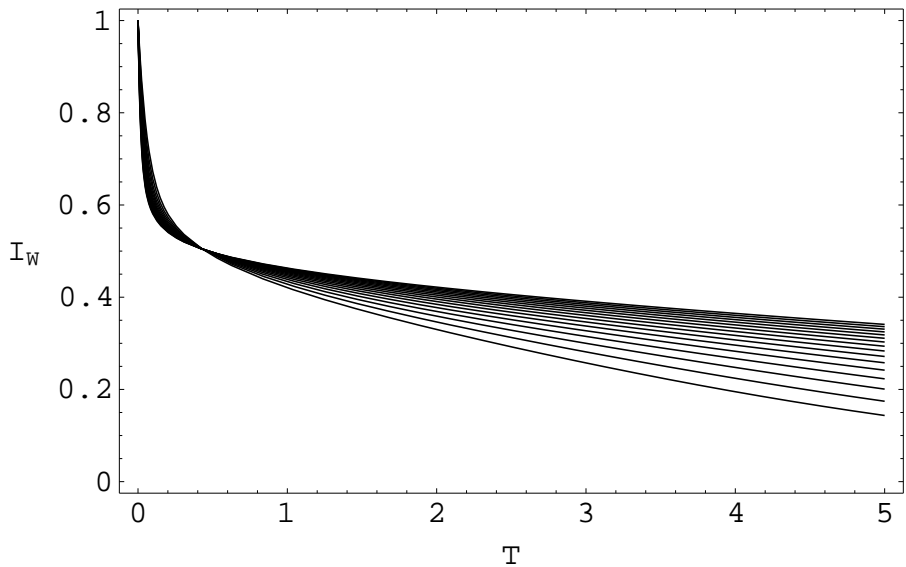}
\caption{ Restricted Witten index for D=2 SYMQM.}
\label{f4}
\end{figure}

Finally let us shortly discuss the Witten index for this model. Because the particle-hole symmetry
interchanges odd and even fermionic numbers, the Witten index vanishes identically. This is also true for any finite
cut-off since \nc\
preserves particle-hole symmetry. Nevertheless one can obtain a non trivial and interesting information
by defining the index {\em restricted} to the one pair of fermion-boson sectors. For example
\eq
I_W(T)^{(0,1)}=\Sigma_{i,F_i=0,1} (-1)^{F_i}\exp{(-T E_i)},
\eqx
where the sum is now restricted only to the $F=0$ and $F=1$ sectors. Since supersymmetry balances
fermionic and bosonic states between these sectors (with the usual exception of the vacuum), the restricted
index is a good and non trivial measure of the amount of the violation/restoration of SUSY, even when
the total Witten index vanishes. Obviously $I_W^{(2,3)}=-I_W^{(0,1)}$, and $I_W^{(0,3)}=I_W^{(1,2)}=0$ .
Studying restricted index
is particularly interesting in this model since, due to the continuum spectrum, it does not have to be integer
and provides some information about the density of states. We have calculated the index from our spectrum of
the original hamiltonian, Eq.(\ref{HD2}), for a range of cut-offs: $ 5 < N_{cut} < 20 $. Figure \ref{f4} shows
the result for the $2+4+4+\dots $ scheme, i.e. when the basis is increased by every four vectors allowing
from the beginning for the two unbalanced states  from the empty and filled sectors each. We see a slow approach
towards the time independent constant which seems to be $1/2$.

Witten index, as a sum over all states, provides an average measure of the breaking/restoration of SUSY. Indeed
even though individual supersymmetric multiplets have not yet clearly formed (at cut-offs used for this calculation),
we see definite flattening of the $T$ dependence  of $I_W(T)$ which must then be a result of the average
cancellation between many levels. Therefore it is easier to see the onset of the supersymmetric
behavior in the Witten index than in the location of the individual levels.
As before Fig. 4 shows that the limiting value of the index is
 discontinuous at $T=0$ with the same interpretation as in Sect.3. However, contrary to the discrete spectrum
 of the Wess-Zumino model, the value of the index is not integer \footnote{It would be an interesting exercise
to calculate the restricted index from the continuous spectrum of Ref.\cite{CH}.}. The value
of the index at $T=0$ is the direct consequence of our scheme ($ 2+4+4+\dots $) of increasing the basis and,
 of course, is scheme dependent. As a final remark we note an intriguing existence of a "universal" point in $T$
 at which the asymptotic value seems to be attained at all \nc.

\section{D=4 supersymmetric SU(2) Yang-Mills quantum mechanics}
\subsection{Preliminaries}
Dimensionally reduced (in space) system is described by nine bosonic coordinates $ x^i_a(t) $,
$i=1,2,3; a=1,2,3$ and six independent fermionic coordinates contained in the Majorana spinor $\psi_a^{\alpha}(t)$,
$\alpha=1,...,4$. Equivalently (in D=4) one could have chosen to impose the Weyl condition and worked with Weyl spinors.
Hamiltonian reads \cite{HS}
\eqn
H=H_B+H_F,   \\ \nonumber
H_B={1\over 2} p_a^ip_a^i+{g^2\over 4}\epsilon_{abc} \epsilon_{ade}x_b^i x_c^j x_d^i x_e^j,\\  \label{HD4}
H_F={i g \over 2} \epsilon_{abc}\psi_a^T\Gamma^k\psi_b x_c^k,
\eqnx
where $\psi^T$ is the transpose of the real Majorana spinor, and $\Gamma$ in D=4 are just the standard Dirac $\alpha$
 matrices. In all explicit calculations we are using Majorana representation of Ref.\cite{IZ}.

Even though the three dimensional space was
reduced to a single point, the system still has the internal spin(3) rotational symmetry,
inherited from the original theory, and generated by the angular momentum
\eq
J^i=\epsilon^{ijk}\left( x^j_a p^k_a-{1\over 4}\psi^T_a\Sigma^{jk}\psi_a\right),\label{JD4}
\eqx
with
\eq
\Sigma^{jk}=-{i\over 4}[\Gamma^j,\Gamma^k].
\eqx
Further, the system has gauge invariance with the generators
\eq
G_a=\epsilon_{abc}\left(x_b^k p_c^k-{i\over 2}\psi^T_b\psi_c \right), \label{GG}
\eqx
and is invariant under the supersymmetry transformations generated by
\eq
Q_{\alpha}=(\Gamma^k\psi_a)_{\alpha}p^k_a + i g \epsilon_{abc}(\Sigma^{jk}\psi_a)_{\alpha}x^j_b x^k_c.
\label{QD4}
\eqx
The bosonic potential (written now in the vector notation in the color space)
\eq
V={g^2\over 4}\Sigma_{jk}(\vec{x}^j\times\vec{x}^k)^2,
\eqx
exhibits the famous flat directions responsible for a rich structure of the spectrum.
At first sight one might expect that the spectrum of the purely bosonic
(hence non supersymmetric) system does not have localized states because of these flat
valleys (c.f. Introduction).
However the flat directions are blocked by the energy of the
transverse quantum fluctuations since valleys narrow as we move from the origin. As a consequence
 the spectrum of the model is discrete. Energies of the first lower states, known as zero
 volume glueballs, were first  calculated by L\"{u}scher and M\"{u}nster \cite{LM}.
On the other hand, in the supersymmetric system, transverse fluctuations cancel among bosons and fermions,
 valleys are not blocked, and the spectrum of the model is expected to be continuous
\cite{dWLN}. The story has its interesting continuation in D=10 model where the
 evidence of the threshold bound state was accumulated \cite{ST}.
Existence of such a state is necessary for the M-theory interpretation of the model where it is
considered as a prototype of the graviton multiplet. Therefore detailed study of the low energy
spectra of the whole
family of Yang-Mills quantum mechanical systems, including identification of the localized and non
localized states is an important and fascinating subject. We shall face some of these questions
also in the D=4 system.
\subsection{Creation and annihilation operators}
There are many ways to write quantum coordinates in terms of creation and annihilation operators
\eq
[a_a^i,a_b^{k\dagger}]=\delta^{ik}_{ab},\;\;\;\{f_a^{\rho},f_b^{\sigma\dagger}\}=\delta_{ik}^{\rho\sigma},
\;\;\rho, \sigma = 1,2 .
\label{aaff}
\eqx

The only constraint comes from the canonical (anti)commutation relations
\eq
[x_a^i,p_b^k]=i\delta^{ik}\delta_{ab} ,\;\; \{\psi_a^{\alpha},\psi_b^{\beta}\}=\delta_{ab}^{\alpha\beta}.
\label{acom}
\eqx
For bosonic variables we just use the standard extension of Eq.(\ref{XPD2}) to more degrees of freedom
\eq
   x^i_a={1\over\sqrt{2}}(a^i_a+a_a^{i \dagger}), \;\;\; p^i_a={1\over i \sqrt{2}}(a^i_a-a_a^{i \dagger}).  \label{XPD4}
\eqx
For fermionic variables we begin with the classical Majorana  fermion in the Weyl representation \cite{WBG}
\eq
\psi_W=(\zeta_2^{\star},-\zeta_1^{\star},\zeta_1,\zeta_2),
\eqx
replace the classical Grassmann variables $\zeta,\zeta^{\star}$ by fermionic creation and annihilation operators
$f,f^{\dagger}$, and transform to the Majorana representation. Final result is a quantum hermitean Majorana spinor
\eq
\psi_a={1+i\over 2\sqrt{2}} \left( \begin{array}{c}
                            -   f_a^{1} - i f_a^{2} + i f_a^{1\dagger} +   f_a^{2\dagger} \\
                            + i f_a^{1} -   f_a^{2} -   f_a^{1\dagger} + i f_a^{2\dagger} \\
                            -   f_a^{1} + i f_a^{2} + i f_a^{1\dagger} -   f_a^{2\dagger}  \\
                            -i  f_a^{1} -   f_a^{2} +   f_a^{1\dagger} + i f_a^{2\dagger}  \\
                                    \end{array} \right),
\eqx
which satisfies Eq.(\ref{acom}) due to (\ref{aaff}). Other choices of fermionic creation and annihilation operators
are also possible \cite{SF,HS,SMK}.

     The next step is to define an empty state and its, computer based, algebraic representation
\begin{eqnarray}
\lefteqn{|(0,0,0),(0,0,0),(0,0,0),(0,0,0),(0,0,0)> \leftrightarrow  } \\
   && \{1,\{1\},\{\{0,0,0\},\{0,0,0\},\{0,0,0\},\{0,0,0\},\{0,0,0\}\}\},\label{D4vac}
\end{eqnarray}
where the first three vectors (in color) specify bosonic, and the last two fermionic, occupation numbers.

 The Jordan-Wigner transformation requires additional specification in this case. As before,
 the action of $f_a^i$ and $f_a^{i\dagger}$ on elementary states is defined by the spin-like raising and
lowering operators corrected by the non local Jordan-Wigner phases
\eq
f_A=\Pi_{B<A}(-1)^{F_B}\sigma_B^{-}\;\;\; f_A^{\dagger}=\Pi_{B<A}(-1)^{F_B}\sigma_A^{+}. \label{JWD4}
\eqx
However, since individual states are now labeled by a {\em double} index $A$, one must define
the ordering of the two dimensional indices. We choose the lexicographic order.
If $A=(a,\alpha)$ and $B=(b,\beta)$ than
\eq
\Pi_{B<A}=\Pi_{\beta<\alpha,b\le 3} \Pi_{\beta=\alpha,b<a} .
\eqx
Any other unambiguous definition of the ordering is admissible.

\subsection{Constructing the basis}
There are many ways to construct bases in higher dimensional systems.
Apart from theoretical requirements one must also take into account practical limitations of computer implementation.
Of course the number of states at finite cut-off is generally bigger for higher D. However, the theory also has
more symmetries and consequently the number of states in a particular channel can be kept manageable.

First important symmetry is the conservation of the number of Majorana fermions
\eq
[F,H]=0\;\;\; F=f_a^{i\dagger}f_a^{i} .
\eqx
That is, the vector interaction $H_F$ cannot produce Majorana pairs in this model. As a consequence the whole Hilbert
space splits into seven sectors of fixed $F=0,1,\dots ,6$.

Second, the system has the particle-hole symmetry
\eq
 F\leftrightarrow 6-F,
\eqx
therefore the first four sectors $F=0,1,2,3$ contain all information about the spectrum.

Further, the gauge invariance requires that our basis is built only from the gauge invariant creators.
All this is not different from the earlier D=2 case.

The new element is the SO(3) invariance, with the angular momentum Eq.(\ref{JD4}), which can be used
to split further the problem into the different channels of fixed $J$. However, more we specify
the basis more complicated its vectors become, and consequently more computer time goes into the generation
of such a basis and subsequent calculation of matrix elements. Similarly one might contemplate using the reduced version of Lorentz invariant composite operators
as possible creators of a basis. Again this would generate more complex basis since fields contain both creation and
annihilation operators. Also, Lorentz covariance is not that relevant in our fixed frame, Hamiltonian formulation.

Taking all above into account we have decided to produce the simplest basis of gauge invariant vectors using
elementary creation operators. To this end we proceed as follows. Consider each fermionic sector separately, i.e.
fix the fermionic number $0 \le F \le 3$.  At given $F$ create all independent vectors with fixed number
of bosonic quanta $B=a_a^{i\dagger} a_a^i$, and define a cut-off as the maximal number of bosonic quanta
$B \le N_{cut}$, independently for each $F$. Hence $N_{cut}$ can depend on $F$. To create all independent, gauge invariant
 states at fixed $F$ and $B$ consider all possible contractions of color indices in a creator of $(F,B)$ order
\eq
a^{i_{1}\dagger}_{a_{1}}...a^{i_{B}\dagger}_{a_{B}}f^{\sigma_{1}\dagger}_{b_{1}}..f^{\sigma_{F}\dagger}_{b_{F}},
\eqx
 for all
values of the spatial indices $i$ and $\sigma$. All color contractions fall naturally into different
{\em gauge invariant classes}. Creators from different classes differ by color contractions {\em between}
 bosonic and fermionic operators. For example
\eq
a^{i\dagger}_a a^{j\dagger}_a a^{k\dagger}_b a^{l\dagger}_b f^{\sigma\dagger}_c f^{\rho\dagger}_c
\eqx
and
\eq
a^{i\dagger}_a a^{j\dagger}_a a^{k\dagger}_b a^{l\dagger}_c f^{\sigma\dagger}_c f^{\rho\dagger}_b
\eqx
belong to different gauge invariant classes according to our definition. Another example involves
odd number of operators where one "contraction" consists of one triple of color indices coupled with the Levi-Civita
symbol \footnote{For the SU(2) gauge group, this step can be generalized to higher SU(N) groups.}. Namely
\eq
\epsilon_{cde}a^{i\dagger}_a a^{j\dagger}_a a^{k\dagger}_b a^{l\dagger}_b a^{m\dagger}_c
                                                                          f^{\sigma\dagger}_d f^{\rho\dagger}_e
\eqx
and
\eq
\epsilon_{cde}a^{i\dagger}_a a^{j\dagger}_a a^{k\dagger}_b a^{l\dagger}_c a^{m\dagger}_d
                                                                          f^{\sigma\dagger}_e f^{\rho\dagger}_b
\eqx
also belong to different gauge invariant classes. Different contractions {\em within} bosonic and fermionic family of
operators do not lead to independent states hence are considered to be in the same class. Still there will be many
linearly dependent states generated by above creators with all values of spatial indices $i$ and $\sigma$. It is
however a simple matter to recognize linearly independent ones given our rules of "quantum algebra". We therefore leave
it to the computer for the time being. At fixed $F$ and $B$ the final procedure is as follows: 1) identify all gauge invariant
classes of creators,
2) loop over all values of spatial indices and for each $i_1,...,i_B,\sigma_1,...,\sigma_F$ create corresponding state
 from the empty state, Eq.(\ref{D4vac}), 3) identify  and reject linearly
dependent vectors, 4) orthonormalize the remaining set of states. In principle this procedure depends exponentially
on $N_{cut}$, hence can be further shortened and improved, however it is sufficiently fast for our present purposes.
It also has an advantage of generating all channels of angular momentum (available within the cut-off \nc).  This will
prove useful in studying supersymmetry.

Table IV shows the sizes of bases we have reached so far in each of the fermionic sectors.
Due to the particle-hole symmetry the structure in the $F=4,5,6$ sectors is identical with that in $F=2,1,0$ respectively.

  \begin{table}
  \begin{center}
   \begin{tabular}{cccccc} \hline\hline
        $F$        &  $ 0 $ & $ 1 $ & $ 2 $ & $ 3 $  & \\  \hline
  $B$ &  $ N_{s}$\ \ \ $  \Sigma$ \ \  & $ N_{s} \ \ \  \Sigma$ \ \  &  $ N_{s} \ \ \  \Sigma$ \ \  & $ N_{s}\ \ \   \Sigma$  \ \ &  B - F \\
   \hline
  0 &    1 \     1 \ \   &  -  \    -   \  &   1 \     1  \  &   4 \     4  \ & 0  \\
  1 &   -  \     1 \ \   &   6 \     6  \  &   9 \    10  \  &   6 \    10  \ & 0  \\
  2 &    6 \     7 \ \   &   6 \    12  \  &  21 \    31  \  &  42 \    52  \ & 0  \\
  3 &    1 \     8 \ \   &  36 \    48  \  &  63 \    94  \  &  56 \   108  \ & 0  \\
  4 &   21 \    29 \ \   &  36 \    84  \  & 111 \   205  \  & 192 \   300  \ & 0  \\
  5 &    6 \    35 \ \   & 126 \   210  \  & 240 \   445  \  & 240 \   540    & 0   \\
  6 &   56 \    91 \ \   & 126 \   336  \  &     \           &                &     \\
  7 &   21 \   112 \ \   &  \    \         &     \           &  \     \       &     \\
  8 &  126 \   238 \ \   &  \    \         &     \           &  \     \       &     \\
\hline
 $j_{max}$ &  8          &  11/2           &     6           &  11/2          &     \\
   \hline\hline
   \end{tabular}
  \end{center}
\caption{Sizes of the bases generated in each fermionic sector, $F$. $N_s$ is the number of basis vectors
  with given number of bosonic quanta, $B$, while $\Sigma$ gives the cumulative size up to $B$. The last
column gives the difference between the total number of the bosonic and fermionic states in all seven sectors.}
\label{bastab}
\end{table}

Calculation of the spectrum of H proceeds exactly as earlier. Naturally we can also readily obtain the spectrum of
other observables, e.g.
of the angular momentum, Eq.(\ref{JD4}), as well as the angular momentum content of the energy eigenstates. For example
the last row of Table 4 gives the highest angular momentum which can be constructed in each sector with
the bases generated so far.

\subsection{Results}

\begin{figure}[htb]
\epsfig{width=12cm,file=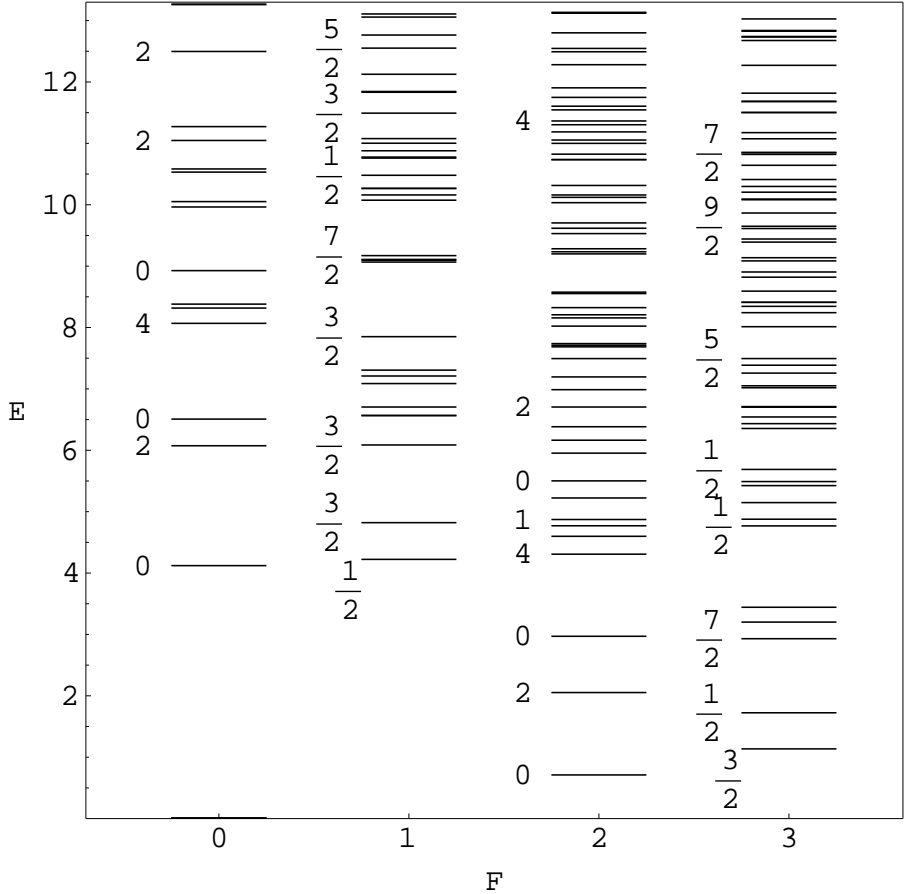}
\caption{Spectrum of D=4 SYMQM.
}
\label{f5}
\end{figure}

The spectrum of the theory is shown in Fig.\ref{f5}. A sample of states is labeled with their angular momenta.
It is important that our cut-off preserves the SO(3) symmetry. Consequently, the basis described
in previous section contains complete
representations of the rotation group for any \nc\ . Hence the spectrum displayed in Fig.5 has appropriate degeneracies
for each value of the angular momentum quantum number $j$.

To maintain some clarity of the figure we have cut arbitrarily
the upper part of the spectrum, which extends to about $E_{max}\sim 35$ with current \nc .
One expects that the individual higher states have a considerable
dependence on \nc.\ However they also carry some relevant information, e.g. for quantum averages.

Apart from relating the bases in corresponding fermionic sectors, particle-hole symmetry also implies
equality of all corresponding energy levels. Since our cut-off respects this symmetry we indeed observe,
in sectors with $F=6,5,4$, exact repetition of the eigenvalues from $F=0,1,2$ subspaces. Therefore only first four
sectors are displayed in Fig.\ref{f5}. Evidently the spectrum of D=4 theory is very rich and raises many
interesting questions. We shall discuss some of them, beginning with the relation to the already
known results.

\subsubsection{Correspondence with pure Yang-Mills quantum mechanics}
\begin{figure}[htb]
\epsfig{width=12cm,file=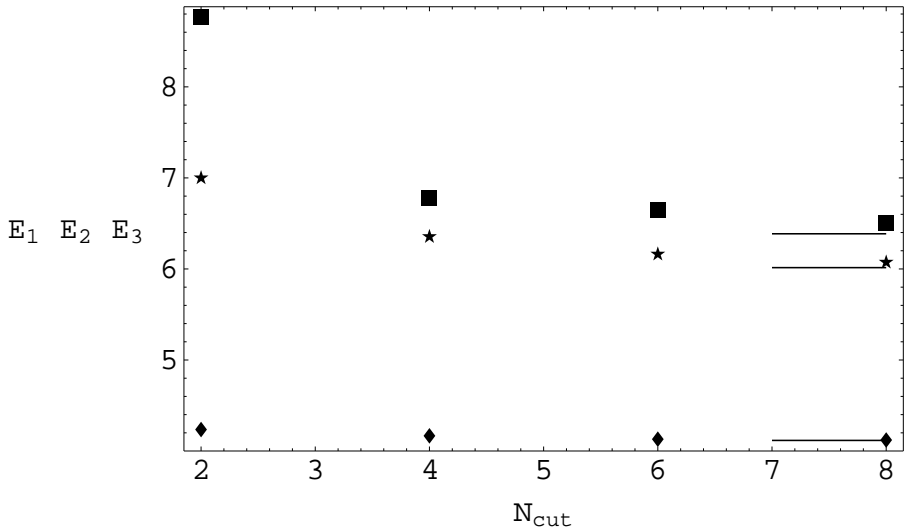}
\caption{ First three energy levels in the $F=0$ sector of D=4 SYMQM for different cut-offs.
Soild lines show the 0-volume results of Ref.\cite{LM}.   }
\label{f6}
\end{figure}

It turns out that the fermionic part of the hamiltonian, Eq.(\ref{HD4}),
annihilates any state in the $F=0$ sector
\eq
H_F |0_F>=0.
\eqx
Therefore the supersymmetric hamiltonian, Eq.(\ref{HD4}) reduces in that sector to the hamiltonian of the
 pure Yang-Mills quantum
mechanics, well known from the small volume approach \cite{L,LM,vBN}. This offers us a chance to compare
one special case of the present spectrum with the classical results  of Ref.\cite{LM}.  Figure \ref{f6}
shows such a comparison for the first three states as a function of \nc.  Nice convergence
of our results towards those of L\"{u}scher and M\"{u}nster is seen. They used 120 states in each angular momentum channel
and variationally optimized frequencies of the basis states.
Our $F=0$ basis contains 238 states which span all angular momenta up to $j=8$, and has only 12 states with $j=0$.
No variational adjustment of the frequencies was attempted.
In view of this we consider above agreement as a rather satisfactory confirmation of the present approach.
The assignment of the angular momentum, which is done automatically within our method, is also the same. In particular,
the characteristic to the 0-volume, ordering of the tensor glueball and the first scalar excitation is reproduced.

\subsubsection{The cut-off dependence}

    The dependence of the spectrum on \nc\ is of course crucial, as it provides the quantitative criterion
if the size of the basis is adequate.
It is displayed in Fig. 7 for the first few  lower levels in all fermionic sectors.
It is clear that the cut-off dependence is different in various fermionic sectors.
For $F=0,1$ (and $F=6,5$) the dependence is rather weak, suggesting that the energies
of lower levels have already converged (within few percent) to their asymptotic values. On the other hand,
in $F=2,3,4$ sectors convergence is considerably slower. As explained earlier, one can use the rate of convergence
with \nc\ to distinguish between the continuous and discrete spectrum. It was proved that in the case of
the free spectrum the convergence is O(1/\nc) i.e. slow \cite{TW}. For the localized bound states
we conjecture that the rate of convergence  with \nc\ is sensitive to the large $x$ asymptotic of the wave function.
 We have observed
this in a simple anharmonic oscillator and other models. This regularity is also clearly confirmed
in the Wess-Zumino and D=2 SYMQM models discussed in previous Sections.

\begin{figure}[htb]
\epsfig{width=14cm,file=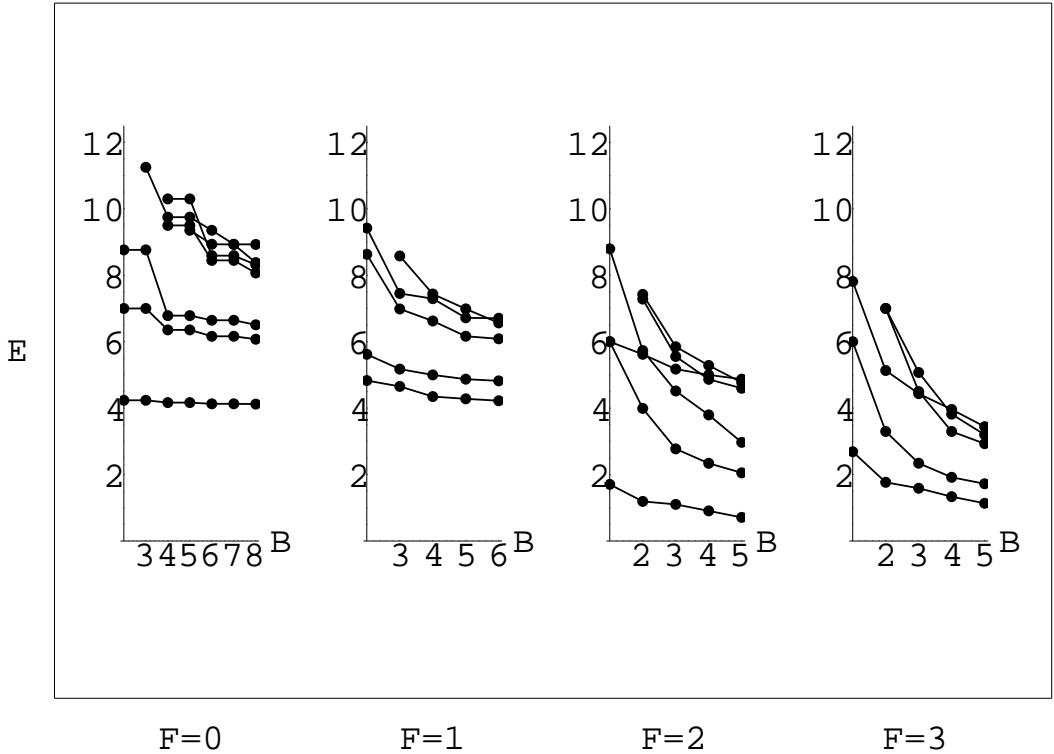}
\caption{ Cut-off dependence of the lower levels of the D=4 SYMQM in four independent fermionic channels.}
\label{f7}
\end{figure}

Taking this into account we claim that the low energy spectrum
of D=4 SYMQM is discrete in $F=0,1,5,6$ and continuous in the $F=2,3,4$ sectors. This is an interesting
quantification of the result of Ref.\cite{dWLN} which was mentioned earlier. Since the fermionic modes are crucial to
provide continuous spectrum, it is natural that it does not show up in the sectors where they do not exist at all,
or are largely freezed out by Pauli blocking.

Second result, which is evident from Fig.7, concerns the supersymmetric vacuum in this model.
Assuming that indeed the eigen energies have approximately converged in $F=0,1$ sectors (none of them to zero)
it follows that the SUSY vacuum can not be in empty and filled sectors ($F=1,5$ is also ruled out by the angular
momentum) \footnote{In early attempts, empty and filled sectors were considered as possible
candidates for SUSY vacuum.}.
The obvious candidates are the lowest states in $F=2,4$ sectors with their energy consistent with zero
at infinite \nc\ , c.f. Fig. 7. It follows, together with the conjectured correlation between localizability and
\nc\ dependence, that the SUSY vacuum in this model is non normalizable.
Further evidence is based on the structure of the supersymmetric multiplets, and will be presented in the next section.
Very recently van Baal studied $F=2$ sector in a more complex,
non compact, supersymmetric model of the same family \cite{vBN}. Although our results cannot be directly compared,
due to the specific boundary conditions required in \cite{vBN}, he finds that the energy of the lowest state in $F=2$
sector is indeed very close to zero. The deviations from exact zero are caused by the SUSY violating boundary
conditions.

It is important to note that not all states in $F=2,3,4$ sectors have to belong to the continuum.
Supersymmetry together with the discrete spectrum in $F=0,1$ channels implies existence of the normalizable
states among the continuum of $F=2,3,4$ states. This will be discussed later, here we only give one explicit example.
Indeed the energy of the $F=2, j=1$ state, shown in Fig. 7 (flatter curve beginning at E=6), has definitely weaker
dependence on \nc\ than the others.
We interpret this as a signature that the lowest $|2_F,1_j>$ state is localized. This situation may be a precursor
of the more complex phenomenon expected in the D=10 theory. There, the zero energy, localized bound state
of $D_0$ branes  should exist at the threshold of the continuous spectrum. Present example suggests that one way to
distinguish such a state from the continuum may be by the different \nc\ dependence.

\subsubsection{Supersymmetry}
{\em Operator level.}
To check the supersymmetry algebra
\eq
\{Q_{\alpha},Q_{\beta}\}=2\delta_{\alpha\beta}H+
2 g \Gamma^k_{\alpha\beta}x^k_a G_a, \label{SALG}
\eqx
at finite cut-off, we sandwich both sides of this operator relation between our basis , introduce the intermediate
complete set of states to saturate the product of $Q$'s and test the equivalent relations between matrix elements
\eq
<N|\{ Q_{\alpha},Q_{\beta} \}|N'>=2\delta_{\alpha\beta}<N|H|N'> .  \label{MSUSY}
\eqx
 The gauge term in Eq.(\ref{SALG}) does not contribute since the basis is gauge invariant
at any \nc . There is, however, one limitation. As is evident from  Eq.(\ref{QD4})), supersymmetry generators change
fermionic number by one, and can change number of bosonic quanta, $B$, at most by two
\eq
<N|Q_{\alpha}|M> \ne 0 \;\;\ if \;\; F_N=F_M\pm 1  , \;\;|B_M-B_N|\le 2 . \label{SULIM}
\eqx
It follows that, if a cut off representation of unity  $ |M><M|,\;\; B_M < N_{cut} $ is used to saturate any matrix
element of a product
$<N|Q|M><M|Q|N'>$, it will be biased for high states in $N,N'$ sectors. Quantitatively, the supersymmetric relations,
Eq.(\ref{MSUSY}) are valid if
\eq
  B_N \;,\; B_{N'} \le N_{cut}-2. \label{SULIM2}
\eqx
This limits the number of matrix elements available for the test. Still many interesting predictions for lower states
can be verified and should be satisfied exactly at finite \nc .

One more property of supersymmetry generators should be kept in mind. Since they change $F$ by 1, they move between
different fermionic sectors. Hence the generic matrix equation (\ref{MSUSY}), when restricted to a particular sector,
 reads ($\alpha=\beta$, no sum)
\begin{eqnarray}
\lefteqn{<F|H|F'>=}\\  \nonumber
&&<F|Q_{\alpha}|F-1><F-1|Q_{\alpha}|F'>+<F|Q_{\alpha}|F+1><F+1|Q_{\alpha}|F'> . \label{SUSYFF}
\end{eqnarray}
That is, the hamiltonian matrix  receives contribution from a single fermionic
sector, only if it is computed in empty or filled sectors. Of course above SUSY relations have many other
consequences. Some of them will be used later, when discussing emergence of the supersymmetric multiplets.

    Within the above limitations we have checked some consequences of SUSY algebra
at finite \nc. The spectrum of the $Q^2$ matrix agrees exactly with that shown in Fig.\ref{f5} and
derived directly form the hamiltonian.
 Moreover, the off-diagonal commutators indeed vanish in our basis. Many other interesting questions
emerge and will be studied elsewhere.

{\em SUSY on the level of states.}
  The next goal is to identify supersymmetric multiplets in the spectrum and to asses the effect
of the cut-off on their splittings and mixings. It is clear from Fig.5 that any direct search for approximately
degenerate states is difficult and may not be conclusive. However our construction provides another simple
way to move
 within the SUSY multiplets. Namely it is sufficient to use the explicit matrix representation of
supersymmetry generators in bases summarized in Table \ref{bastab}.
If supersymmetry was exact, acting with $Q$ on an eigenstate of $H$, would
generate states within the same multiplet. At finite \nc\ this is no longer true, however we expect that
the resulting state, $Q|\Psi>$ say , will be spread around the appropriate multiplet member (or members)
in another fermionic sector. This information will then be correlated with the spectrum of Fig. 5.
Important simplification comes from the rotational symmetry, which is exact at every \nc\  in this approach.
Supersymmetric generators carry the angular momentum 1/2 which clearly limits the range of possible targets
$Q|\Psi>$. Indeed, we confirm that
\eq
<F,j|Q|F\pm 1,j'>=0\;\; if \;\;|j-j'| \ne {1\over 2} .
\eqx

\begin{figure}[htb]
\epsfig{width=12cm,file=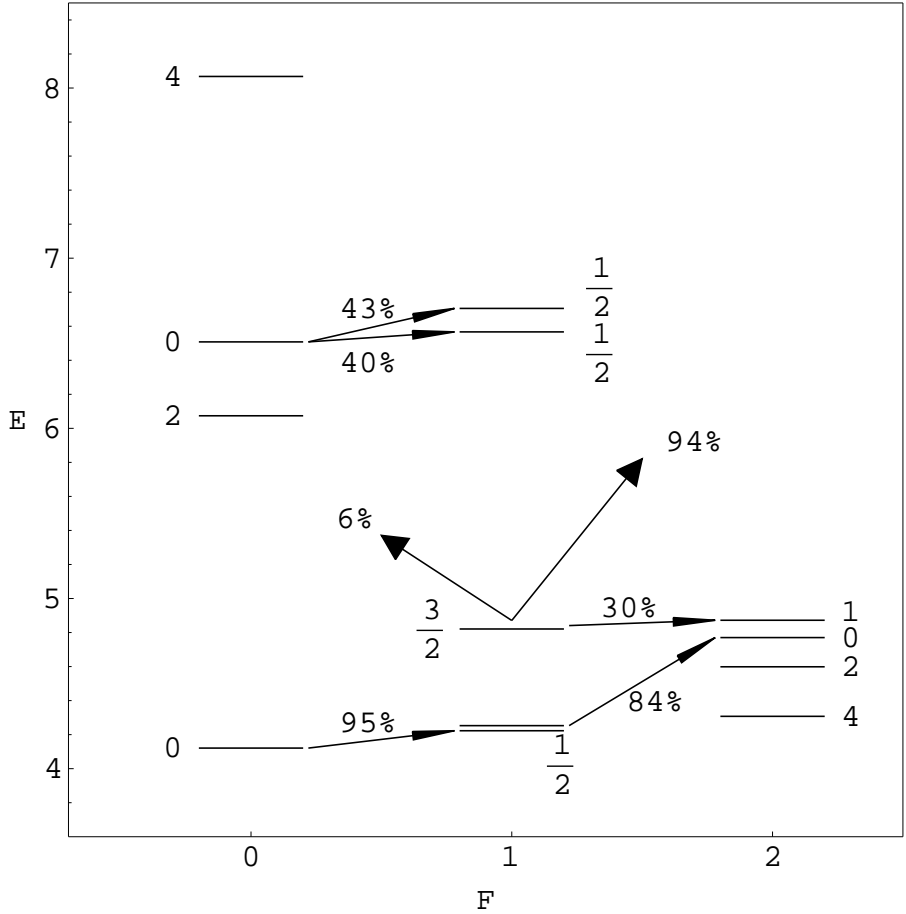}
\caption{ Supersymmeric images of a sample of eigenstates.}
\label{fig:f8}
\end{figure}

 To proceed, we  denote the $k$-th energy eigenvector in the
$m$-th fermionic sector and the $n$-th angular momentum channel  by $|m_F,n_j;k>$.
The subscript $j$ will be omitted where evident. A summary of various transitions we have analyzed is shown in
Fig.\ref{fig:f8}. For example,  the action of $Q_1$ on the first state in $F=0$ sector
gives
\eq
{(|<1_F,1/2;1|Q_1|0_F,0_j;1>|^2+|<1_F,1/2;2|Q_1|0_F,0_j;1>|^2) \over ||Q_1|0_F,0_j;1>||^2}=95\%,
\eqx
which provides rather satisfactory confirmation of the first-glance guess from Fig.5. For higher states
situation is more complex.  The second scalar state in this sector goes under $Q_1$ into
$Q_1|0_F,0_j;2>$ which decomposes in 40\% into the $|1_F,1/2;3,4>$ doublet and 43\% of yet higher,
 but also close in energy,
$|1_F,1/2;5,6>$ doublet. This is not surprising, since higher states have
not yet converged to their infinite \nc\ limit.

Even more interesting is to analyze SUSY relations between $F=1$ and $F=2$ sectors. Again the lowest
 $|1_F,1/2;1>$ state transforms predominantly (84\%) into the third scalar $|2_F,0_j;3>$ which is closest in the
energy. Another consequence of the SUSY algebra can be seen here. Naively one would expect
that combining above transitions $|0_F,0_j;1> \rightarrow |1_F,1/2;1>$ and $|1_F,1/2;1>\rightarrow|2_F,0_j;3>$
would lead to the connection of $F=0$ and $F=2$ sectors and extension of the multiplet. This contradicts Eq.(\ref{MSUSY})
which implies that the $Q_1^2$ is diagonal in the eigen basis of the hamiltonian. Solution of this apparent paradox
is simple.
There are two states with $j=1/2$ and while the multiplet beginning in $F=0$ sector ends on one linear combination,
the second multiplet must start on the orthogonal one. In other words the supersymmetric images of $|0_F,0_j;1>$
and $|2_F,0_j;3>$ in $F=1$ sector must be orthogonal. We confirm this exactly
(e.g. to the full precision of Mathematica numerics)\footnote{This is because we use the limited basis Eq.(\ref{SULIM2}),
which guarantees Eq.(\ref{MSUSY}) exactly.}
\eq
<2_F,0_j|Q_1^{\dagger} Q_1|0_F,0_j> \sim 10^{-16}.
\eqx
The next (in the energy)
 $|1_F,3/2;1>$ state goes under $Q_1$ in 30\% to $|2_F,1_j;1,2,3>$ triplet, with the rest shared among higher $F=2$
states.
In fact, the $|2_F,1_j;1,2,3>$ states are those with the fast \nc\ convergence, which we have already pointed earlier
as good candidates for localized states.
Very little of $|1_F,3/2;1>$ goes back to the $F=0$ sector (c.f. fat head arrows in Fig.\ref{fig:f8})
\eq
{||Q_1|1_F,3/2;1>||_{F=0}^2 \over ||Q_1|1_F,3/2;1>||^2 }=6.3\% ,
\eqx
since there is no state with similar energy there \footnote{All surrounding states have already converged within
few percent which suggests that the above small rate is also a O(1/\nc) effect.}.
This means that the $j=3/2$ states start a new SUSY multiplet
which extends into $F=2$ sector and not into $F=0$ one.

We are using only one of the four SUSY generators just as an example. Similarly we do not consider systematically
all members of $j\ne 0$ multiplets. The full analysis, which in fact would be quite straightforward in the present setup,
 would involve complete discussion of the extended, $N=4$, SUSY algebra. However it is out of the scope of this
article and will be done elsewhere.

All above examples have a common feature. Clear SUSY regularities show up if the states
have already converged, i.e. if their energy depends weakly on \nc\ .
This is not the case for the lowest states in $F=2,3,4$ sectors which, we believe, form a continuum at infinite cut-off.
Accordingly, and similarly to the D=2 SYM QM case, identification of multiplets and the SUSY vacuum in this sector
is not clear at the moment. As one possible signature of the vacuum one might take the average norm of
the supersymmetric image of the lowest state $||Q_1|2_F,0_j;1>||^2/dim[3_F]$. Indeed it is smaller than that
for other states and falls with \nc. However it is too early to draw definite conclusions.

Clearly we have only begun. The aim here is just to introduce the new approach and its potential \footnote{We know of no
other method capable to provide such a detailed and quantitative characteristics as discussed in this Section.}, leaving
specific applications for more focused articles.

 {\em Witten index.} The total number of bosonic and fermionic states is the same for each $B$,
c.f. last column of Table 4.
This is not a direct consequence of supersymmetry, but rather of a combinatorics of binary fermionic systems, which
is not spoiled by gauge invariance.
Total number of states gives the Witten index at $T=0$. Hence we expect
\eq
\lim_{N_{cut}\rightarrow\infty} I_W(0,N_{cut})=0.
\eqx
As was already discussed Witten index is discontinuous at $T=0$. In particular $I_W(0)$ depends on the
regularization, therefore this number is tied to our B - F symmetric scheme of increasing the basis.

\begin{figure}[htb]
\epsfig{width=12cm,file=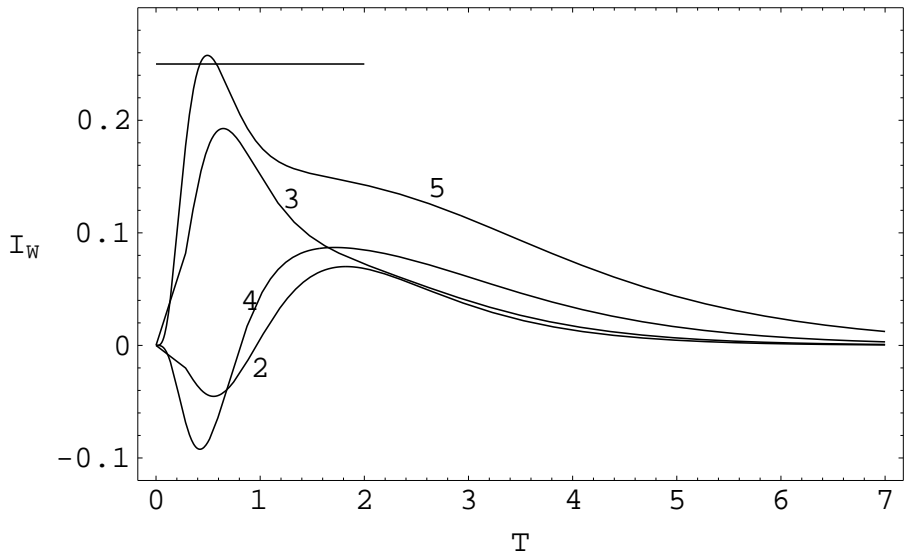}
\caption{ Witten index of the D=4, supersymmetric Yang-Mills quantum mechanics for the number of bosonic quanta
bounded by $N_{cut}=2,...,5$. The bulk value 1/4 is also shown. }
\label{fig:f9}
\end{figure}

Figure 9 shows the (euclidean) time dependence of the Witten index calculated from our spectrum with up to five
bosonic quanta.
Clearly we are far from the convergence, but some interesting signatures can be already observed.
First, one can see how the discontinuity at $T=0$ emerges even at this early stage. Second, the exponential
fall-off common to all curves at large $( T > 5 )$ times is evident. This is where the lowest state dominates.
Since at this \nc\ its energy is not zero, the fall-off is exponential. Finally, and most interestingly,
we observe the flattening shoulder appearing at $T\sim 2-3$. This is the signal of the supersymmetric cancellations
which occur on the average, even though the exact SUSY correspondence between individual states does not yet appear.
The behavior with \nc\ is not inconsistent with the exact bulk value 1/4 \cite{ST} obtained also from the
nonabelian integrals \cite{GG,SM,NSM}. Clearly higher \nc\ are needed, and, what is important, they are perfectly within
 the range of present computers.

\section{Summary}

The new approach to quantum mechanical problems is proposed. The Hilbert space of quantum states is algebraically
implemented in the computer code. In particular realization, used here, states are represented as Mathematica lists.
All basic operations on quantum states are mirrored as definite operations on above lists.
Any quantum observable is represented as a well defined function on these lists. This allows for automatic
calculation of matrix representation of a hamiltonian and other quantum operators of interest. To this end we use
the discrete eigen basis of the operators of the numbers of quanta. The length of above lists is not fixed.
Similarly to the length of an arbitrary quantum combination of basic states, it can vary dynamically allowing for
 any number of quanta.

Of course in any finite computer the maximal length of a combination must be limited. Therefore we impose a cut-off \nc\
which bounds from above the number of allowed quanta. The stringent and quantitative test of this approach is provided by
checking the dependence of any physical observable, e.g. the spectrum or the wave functions, on the cut-off. This is similar
to studying the infinite volume statistical systems via the finite size scaling. We have applied
above technique to the three progressively more complicated systems: Wess-Zumino quantum mechanics, supersymmetric Yang-Mills
quantum mechanics in D=1+1 dimensions, and D=4 SYM quantum mechanics, both for the SU(2) gauge group. In distinction
from many other approaches (for example lattice simulations) the method is completely insensitive to the sign
problem and works equally well for bosonic and fermionic systems.

For Wess-Zumino quantum mechanics we can calculate the discrete spectrum for any values of parameters. Clear restoration
of the supersymmetry was observed for the cut-offs well within the capabilities of present computers. Witten index was
also obtained and its convergence to the known result $I_W(T)=2$ is clearly seen.

The next system, D=2 SYM QM, possesses the gauge invariance which was readily incorporated in our approach.
The physical subspace of gauge invariant states was explicitly constructed.  Known structure
of the solutions in terms of four fermionic sectors was reproduced. Convergence of the method,
and emergence of the supersymmetry,
was also studied in this more difficult case of the continuum spectrum. Witten index, restricted to one
supersymmetric branch of the model, was defined and computed for the first time. Clear, albeit slow, convergence
to the time independent fractional value $I_W^R(T)=\pm 1/2 $ was observed. Moreover, we have found a special scheme
of increasing the basis such that the supersymmetry is {\em exact} at any finite cut-off. This however, may be related
to the exact solubility of the model.

Finally, the method was applied to the unsolved up to now D=4 SYMQM. This much richer system has the SO(3)
rotational symmetry inherited from its space extended predecessor. Our approach preserves this symmetry exactly
at any finite cut-off. The Hilbert space splits again into seven sectors with fixed fermionic number.
We have obtained the complete spectra in all these sectors and studied their cut-off dependence. The spectrum in $F=0$ sector
agrees with the classical 0-volume calculation for pure Yang-Mills quantum mechanics \cite{LM}. An efficient method
to distinguish between the discrete and continuous spectrum was proposed. It turns out that the asymptotics of the
wave function at large distances determines the convergence of our calculations with the number of allowed quanta \nc.
The continuous spectrum with non localized wave functions converges slowly $(\sim O(1/N_{cut})$, while discrete,
localized bound states lead to faster (sometimes even exponential) convergence. Accordingly, we have found an evidence
that
the spectrum of D=4 SYM QM is discrete in $F=0,1,5,6$ sectors while it is continuous in the $F=2,3,4$ sectors.
This provides
an explicit realization of the claim of Ref.\cite{dWLN} in particular fermionic sectors. Interestingly, localized states
exist also in the sectors with continuous spectrum. This is a simple consequence of supersymmetry whose transformations
move between adjacent fermionic sectors. Our method allows to monitor directly the action of SUSY generators and analyse
supersymmetric images of any state. In this way we have identified some candidates for lowest supersymmetric multiplets.
They do not have the same energy at current values of the cut-off, however the splittings are small and consistent
with vanishing at infinite \nc. In particular, the 0-volume glueballs found in $F=0$ sector are relevant to fully
supersymmetric theory in that there exist gluino-gluon bound states with the same masses. We have also
found a candidate for the supersymmetric vacuum which seems to belong to the continuum. However identification of
SUSY multiplets in the continuum part of the spectrum requires higher \nc.  Supersymmetry was also tested
on the operator level.
In particular we confirmed that the spectrum of the hamiltonian coincides with that of $Q_{1}^2$.
As a final application we have calculated the Witten index for this theory. It still depends strongly on
 \nc\ . Nevertheless an early evidence for some of its asymptotic properties can be already seen.

\section{Future prospects}
The main goal of this work is to asses the feasibility of attacking with the new approach the BFSS model of M-theory.
Certainly the method is more efficient for smaller number of degrees of freedom. Application to the Wess-Zumino
quantum mechanics and D=2 Yang-Mills quantum mechanics give quite satisfactory and quantitative results including
the new intriguing scheme which preserves exact supersymmetry for finite cut-off. For more complex, and correspondingly
richer, D=4 SYMQM the method is able to provide new results including detailed information about the supersymmetric
structure of the spectrum and the observables. Obviously we also see a room for improvement which is especially needed
in the continuum sector.
However, and that we find most important, further increase of the size of the Fock space is possible
within the available technology. The whole programme can be (and is being) implemented in the standard, compiler based
languages which usually improves performance by a factor 10-100. Further, the action of the quantum operators
can be optimized taking into account symmetries of the states. Finally, one can go for more powerful computers.

Taking all above into account, one can reasonably expect substantial improvement in the quality of the present
D=4, N=2 results. At the same time one should be able to study D=4 systems with higher N.
To answer the main question: yes, we think that the quantitative study of the $D=10$ theory is feasible, and reaching
current quality results for the D=10 is realistic, beginning with the SU(2) gauge group. $D=10$ hamiltonian and $spin(9)$
generators do not conserve fermionic number \cite{HS,SMK}. The reason for this complication should be better
understood in the first place.

Apart from increasing the number
of degrees of freedom, higher gauge groups pose an interesting problem of constructing gauge invariant states.
This has already been done in the small volume approach for SU(3) \cite{PW}, and should not present any fundamental
problem for higher N as well. At the same time the possibility of some large N simplification, specifically
within the present approach, should be investigated.

Last, but not least, we would like to mention a host of applications of this method to other quantum mechanical systems.
For example one may simply extend the 0-volume calculations to the full non supersymmetric QCD with dynamical quarks
in fundamental representation. Apart from known glueballs, this would give us a spectrum of quark-made-hadrons in
the "femtouniverse".

As another rather different application, we mention that this program is already being used as a routine in solving,
to a high precision, quantum mechanics of a simple
two dimensional building blocks of a prototype quantum computer \cite{LEO}. In particular, a complete quantum evolution
in time with the time dependent hamiltonian, was straightforward to simulate.

The D=4, N=2 SYMQM studied here has 15 degrees
of freedom. There are many unsolved quantum mechanical systems with this or smaller complexity. Present approach
should be well applicable in some of these cases.

Obviously there are many routes which can be followed from this point and we are looking forward to explore some of them.

\vspace*{0.5cm}

\noindent {\em Acknowledgments.} I would like to thank C. M. Bender for an instructive discussion which
inspired this approach. I also thank P. van Baal, P. Breitenlohner, L. Hadasz,  M. Rostworowski and H. Saller
 for intensive discussions. This work is  supported by the
Polish Committee for Scientific Research under the grant no. PB 2P03B01917.


\begin{thebibliography}{99}
\bibitem{BFSS} T. Banks, W. Fishler, S. Shenker and L. Susskind,
 Phys.Rev. {\bf D55} (1997) 6189; hep-th/9610043.
\bibitem{WI} E. Witten, Nucl. Phys. {\bf B185/188} (1981) 513.
\bibitem{CO} F. Cooper, A. Khare and U. Sukhatme, Phys. Rep.
 {\bf 251}(1995) 267, hep-th/9405029.
\bibitem{CH} M. Claudson and M. B. Halpern, Nucl. Phys. {\bf B250} (1985) 689.
\bibitem{STS} S. Samuel, Phys. Lett. {\bf B411} (1997) 268, hep-th/9705167.
\bibitem{SF} U. H. Danielsson, G. Ferretti and B. Sundborg,
Int. J. Mod. Phys. {\bf A11} (1996) 5463, hep-th/9603081.
\bibitem{HS} M. B. Halpern and C. Schwartz, Int. J. Mod. Phys. {\bf A13} (1998) 4367, hep-th/9712133.
\bibitem{dWLN} B. de Wit, M. L\"{u}scher and H. Nicolai, Nucl. Phys. {\bf B320} (1989) 135.
\bibitem{NH} H. Nicolai and R. Helling,
In {\em Trieste 1998, Nonperturbative aspects of strings, branes
and supersymmetry} pp 29-74. , hep-th/9809103.
\bibitem{L} M. L\"{u}scher, Nucl. Phys. {\bf B219} (1983) 233.
\bibitem{POL} J. Polchinski, {\em String Theory}, Cambridge University Press, Cambridge, 1998.
\bibitem{YI} P. Yi, Nucl. Phys. {\bf B505} (1997) 307, hep-th/9704098.
\bibitem{ST} S. Sethi and M. Stern, Comm. Math. Phys., {\bf 194} (1998) 675, hep-th/9705046.
\bibitem{SM} A. V. Smilga, Nucl. Phys. {\bf B266} (1986) 45.
\bibitem{SMK} V. G. Kac and A. V. Smilga, Nucl.Phys. {\bf B571} (2000) 515, hep-th/9908096.
\bibitem{NSM} G. Moore, N. Nekrasov and S. Shatashvili,
Commun. Math. Phys.{\bf 209} (2000) 77,  hep-th/9803265.
\bibitem{GG} M.B. Green and M. Gutperle, JHEP {\bf 01} (1998) 005,hep-th/9711107.
\bibitem{SU} F. Sugino, Int. J. Mod. Phys. {\bf A14} (1999) 3979, hep-th/9904122.
\bibitem{KS} W. Krauth, H. Nicolai and M. Staudacher, Phys. Lett. {\bf B 431} (1998) 31,
hep-th/9803117;
 \bibitem{KSII} W. Krauth and M. Staudacher, Nucl.
Phys. {\bf B584}(2000) 641, hep-th/0004076.
\bibitem{KAB1} D. Kabat and G. Lifschytz, Nucl.Phys. {\bf B571} (2000) 419, hep-th/9910001.
\bibitem{KAB3} D. Kabat, G. Lifschytz and D. A. Lowe, Phys. Rev. {\bf D64} (2001)
124015 , hep-th/0105171.
\bibitem{MAR} E. Martinec, in {\em Cargesse 1999, Progress in
string theory and brane theory}, pp 117-145, hep-th/9909049.
\bibitem{JW} R. A. Janik and J. Wosiek, Acta Phys. Polon {\bf B32 } (2001) 2143,
hep-th/9903121.
\bibitem{BW} P. Bialas and J. Wosiek, Nucl. Phys. {\bf B} (Proc. Suppl.) {\bf 106}
 (2002) 968 , hep-lat/0111034.
\bibitem{EK} T. Eguchi and H. Kawai, Phys. Rev. Lett. {\bf 48}
(1982) 1063.
\bibitem{TG} A. Gonzalez-Arroyo, J. Jurkiewicz, and C. P.
Khortals Altes, in {\em Proc. 11th NATO Summer Institute}, eds. J.
Honercamp et al. (Plenum, New York, 1982).
\bibitem{IKKT} N. Ishibashi, H. Kawai, Y. Kitazawa, and A.
Tsushijya, Nucl. Phys. {\bf B498} (1997) 467.
\bibitem{IKAMB} H. Aoki et al., Prog. Theor. Phys. Suppl. {\bf 134} (1999) 47,
 hep-th/9908038.
\bibitem{AMB} J. Ambjorn et al., JHEP 0007 (2000) 011 hep-th/0005147.
\bibitem{AMBII} J. Ambjorn, K.N. Anagnostopoulos and
A. Krasnitz, JHEP 0106 (2001) 069 hep-ph/0101309.
\bibitem{LM} M. L\"{u}scher and G. M\"{u}nster, Nucl. Phys. {\bf B232} (1984) 445.
\bibitem{VABA} P. van Baal, Acta Phys. Polon. {\bf B20} (1989) 295.
\bibitem{vBS} P. van Baal, in {\em At the Frontiers of Particle Physics - Handbook of QCD, Boris Ioffe Festschrift},
vol. 2, ed. M. Shifman (World Scientific, Singapore 2001) p.683; hep-ph/0008206.
\bibitem{vBN} P. van Baal, hep-th/0112072, to appaear in the Michael Marinov Memorial
 Volume {\em Multiple Facets of Quantization and Supersymmetry}, ed. M. Olshanetsky and
 A. Vainstein, World Scientific.
\bibitem{CMB1} C. M. Bender et al., Phys. Rev. {\bf D32} (1985) 1476.
\bibitem{CMB2} C. M. Bender and K. A. Milton, Phys. Rev. {\bf D34} (1986) 3149.
\bibitem{SHIF} M. A. Shifman, {\em ITEP Lectures on Particle Physics and Field Theory}, World Scientific, Singapore, 1999.
\bibitem{JOWI}  P. Jordan and E. P. Wigner, Z. Phys. {\bf 47} (1928) 631.
\bibitem{BD} J. D. Bjorken and S. D. Drell, {\em Relativistic Quantum Fields} v.2, McGraw-Hill, Inc., New York, 1965.
\bibitem{FromR}  A. Nakamura and F. Palumbo, Phys. Lett. {\bf B135} (1984) 96.
\bibitem{TW} J. Trzetrzelewski and J. Wosiek, in preparation.
\bibitem{IZ} C. Itzykson and J.-B. Zuber, {\em Quantum Field Theory}, McGraw-Hill, New York, 1980.
\bibitem{WBG} S. Weinberg, {\em The Quantum Theory of Fields III - Supersymmetry}, Cambridge University Press,
Cambridge, 2000.
\bibitem{PW}  P. Weisz and V. Ziemann, Nucl. Phys. {\bf B284} (1987) 157.
\bibitem{LEO} V. Corato et al., in preparation.







\end{thebibliography}
\end{document}